\documentclass[5p,twocolumn,authoryear]{elsarticle}
\usepackage[latin2]{inputenc}         

\journal{Journal of \LaTeX\ Templates}

\usepackage{url}
\usepackage{array}
\begin{document}

\begin{frontmatter}

\title{Probability of coincidental similarity among the orbits of small bodies -- I.  Pairing}


\author[mymainaddress]{Tadeusz Jan Jopek\corref{mycorrespondingauthor}}
\cortext[mycorrespondingauthor]{Corresponding author}
\ead{jopek@amu.edu.pl}

\author[mysecondaryaddress]{Małgorzata Bronikowska}

\address[mymainaddress]{Institute Astronomical Observatory, Faculty of Physics, A.M. University, Pozna\'{n}, Poland}
\address[mysecondaryaddress]{Institute of Geology, Adam Mickiewicz University, ul. Krygowskiego 12, 61-606 Pozna\'{n}, Poland}

\begin{abstract}
Probability of coincidental clustering among the orbits of comets, asteroids and meteoroids depends on many factors like: the size of the orbital sample searched for clusters or the size of the identified group, it is different for groups of 2,3,4, ... members. Probability of coincidental clustering is assessed by the numerical simulation, therefore, it depends also on the method used for the synthetic orbits generation.

We tested the impact of some of these factors.
For a given size of the orbital sample we have assessed probability of random pairing among several orbital populations of different sizes. We have found how these probabilities vary with the size of the orbital samples. 

Finally, keeping fixed size of the orbital sample  we have shown that the probability of random pairing can be significantly different for the orbital samples obtained by different observation techniques. 
Also for the user convenience we have obtained several formulae which, for given size of the orbital sample, one can use to calculate the similarity threshold corresponding  to the small value of the probability of coincidental similarity among two orbits.   
\end{abstract}


\end{frontmatter}
%
\section{Introduction}
Determining a pairing among small bodies of the Solar System attracted some interest recently. \cite{VokNes2008, Rozek2011} searched for Main Belt asteroid pairs suspected of having a common
origin. \citet{Spiny} has found that two chondrite meteorites, P\v{r}ibram observed in April 1959 and  Neuschwanstein observed in April 2002, moved on a very similar orbits. The authors claim that a very strong resemblance of these orbits argue the common origin of these bodies. \citet{Ohtsuka} searched for meteoroid orbit similar to the Apollo-type near-Earth Asteroid (25143) Itokawa. 
  
The major tool for determining a pairing among small bodies of the Solar System has been orbit similarity, quantified by a function called D-criterion. A number of D-criteria have been developed, the first by \citet{SouthworthHawkins} and its variations by  \cite{ Drummond, Steel91, Jopek1993, Valsecchi1999, Jopek2008, Rudawska2014}. 

Finding a very similar orbits among the asteroids, comets or meteoroids always  rise a  question --- whether such similarity is only a chance coincidence? 
To answer this question we need adequate value of the orbital similarity threshold (a key parameter of any cluster analysis) corresponding to some fixed probability level of a chance similarity between two orbits. 

The same problem arises when we search for meteoroid streams or asteroid families  --- one can inquire about the level of the statistical significance of the groups found by some cluster analysis method applied with a given orbital sample. 

Reliability of orbital grouping  is quite old problem, first time engaged by \citet{SouthworthHawkins} and \citet{Nilsson1964}. The authors investigated the reliability of meteoroid streams found among observed photographic and radio meteor orbits. For this purpose they used equivalent sets of artificial data that were constructed by shuffling and re-assigning appropriate meteoroid orbital elements.
Nilsson claims that the number of spurious streams found in such artificial samples is an upper limit to that in the true sample.
Due to limitation of the computing power at that time it was rather problematic to
accomplish an extensive reliability test of detected meteoroid streams. So no probability values were assigned to the identified pairs or group of meteoroid orbits. 
Instead, using a four-dimensional point distribution as a model of distribution of
meteoroid orbits \citet{SouthworthHawkins} concluded that the orbital similarity threshold $D_{c}$ should vary inversely to the fourth root of the orbital sample size~$N$. Following \citet{Lindblad2} we have the formula:
\begin{equation}
D_{c}=0.2\left(\frac{360}{N}\right)^{1/4}\approx 0.8712 N^{-1/4},
\label{row:01}
\end{equation}
The formula was calibrated with $N=359$ of photographic meteors observed by the Super-Schmidt cameras. The values of $D_C$ given by (\ref{row:01}) as well as its slight variety \citep{Lindblad2} have been applied for identification of meteoroid streams, i.a. in: \citet{SouthworthHawkins, Lindblad1, Lindblad2, Lindblad3, Lindblad4, Jopek1986} and quite recently by \citet{Holman2012}.


However, in case of the different technique, e.g. the meteors observed by small photographic cameras or meteors observed by radar systems, application of this formula may prove to be risky. Therefore in practice the authors in question, and also \citet{GartrellElford, Jopek1986} carried out several meteoroid stream searches at different values of $D_c$ close to the value given by formula (\ref{row:01}). 

In terms of probability,  estimation of the reliability of the identified meteoroid streams was done by \citet{Jopek1997}. For given value of $D_c$ the probability of finding a spurious group of meteoroids of $M$ members was assessed by the cluster analysis in the  two hundred artificial orbital samples.    
In \citet{PaulsGladman},  to asses such probability the authors proposed interesting approach based on the geometric probability distribution of a series of Bernoulli trials. The authors developed a method for calculating the probability of random pairing among two orbits in a given orbital data set.

In the present study, using the ideas described in \citet{Jopek1997, Jopek2003} and in  \cite{PaulsGladman}, we show how different factors can influence the assessment of the probability of chance grouping among the orbits taken from different data samples: bolides, video, and  radio meteors. Also, we have scoped the probability of spurious pairing among the near Earth asteroids (NEAs).
\section{Generation of the orbital samples}
Our study is based heavily on the artificial orbital samples derived by means of the observed orbits. The synthetic orbits free from clusters of common origin one can derive by different methods. Therefore we have investigated how the choice of such method impacts the assessment of the probability of the spurious pairing among the small bodies orbits?
\subsection{Data sources}
We used a part of the meteor data set as in the paper by \citet{Jopek2013}. A subset of  $32205$  sporadic meteoroids were taken from:
\begin{itemize}
\item the IAU MDC database (photographic meteors: Super Schmidt, small camera; radio meteor data: Harvard,  Kharkov,) see \cite{LindbladSteel, Lindblad2003, Svoren2008, Porubcan2011},
\item  photographic and video meteors observed by members of Dutch Meteor Society (DMS) (Betlem {\it et al.} 1998, 2000),
\item  video meteors observed by Japanese amateur astronomers in year 2009 --- SonotaCo group (SonotaCo 2011).
\end{itemize}
\begin{table}
\footnotesize
\caption[] {Sources of meteoroids and NEAs orbits used in this study. $32205$ sporadic meteoroid orbits were selected from several sources;
the NEAs orbits were taken  from the NEODyS website.}
\label{ta:01}
\begin{center}
\begin{tabular}{|r|l|l|}
\hline
\multicolumn{1}{|c|}{ Sample} & \multicolumn{1}{c|}{Sample} & \multicolumn{1}{c|}{}\\
\multicolumn{1}{|c|}{ size} & \multicolumn{1}{c|}{type} & \multicolumn{1}{c|}{}\\
\hline
 495  &  bolides      & meteors brighter than $-4.5^m$\\
1098  &  photo.      & Super-Schmidt, small camera data\\
13968 &  radio        & Harvard (B) synoptic year sample\\
 4136 & radio         & Kharkov sample \\
12508 &  video        & SonotaCo, 2009 sample \\
12057 &  NEAs         & NEODyS-2 Jan. 2015 sample \\
\hline
\end {tabular}
\end {center}
\normalsize
\end {table}
\begin{figure}
\centerline{
\includegraphics[width=0.23\textwidth]{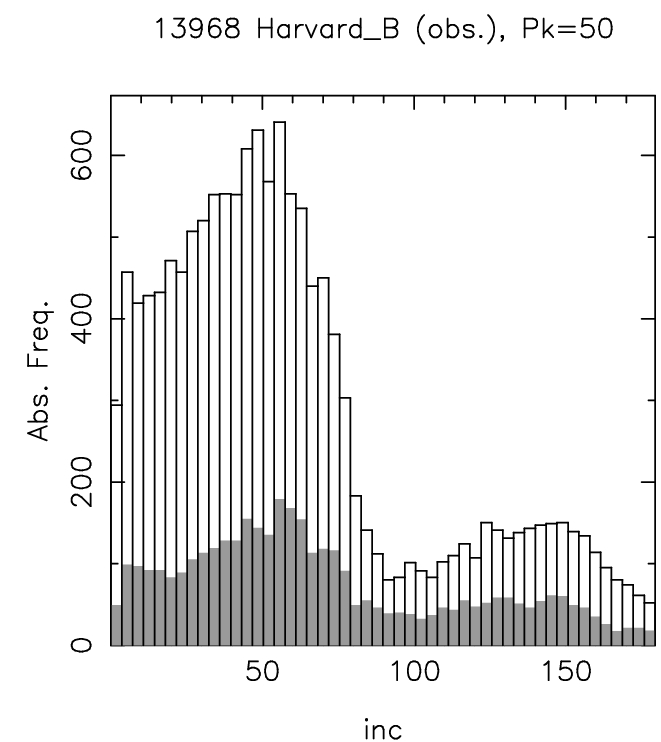}
\includegraphics[width=0.23\textwidth]{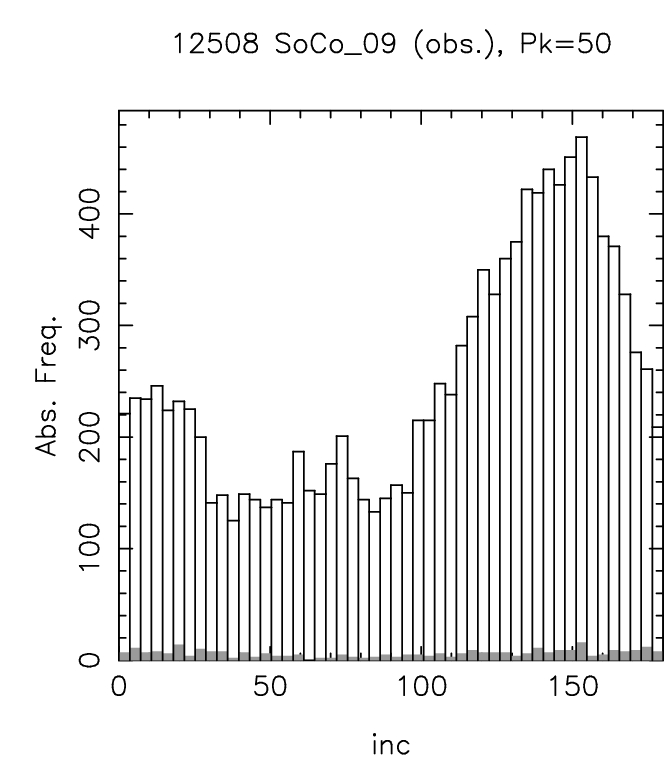}
}
\caption{Histograms of the orbital inclinations of sporadic meteoroids observed by radio (on the left) and video (on the right) techniques.
The histograms  differ significantly. For Harvard (B) sample, $19.9$\% of meteoroids moved on the retrograde orbits, for SonotaCo (2009) $64.9$\% respectively. 
The shadowed beans correspond to the orbits with semi-major axes $a < 1.0$ [AU]. In Harvard sample we have $27$\% such orbits, in SonotaCo sample only $2.5$\% orbits.
}
\label{fig:01}
\end{figure}
The sample of $\sim$$32000$ sporadic meteoroid orbits (see Table \ref{ta:01}) incorporates data obtained by various techniques. 
In addition we used the subset of the NEAs orbits taken from the \citet{NeoDys}  website. From the original data set, $16$ NEA orbits were rejected: all for which $i>76\deg$ or $q>1.3$ [AU], or $a>5.0$ [AU]. As result our NEAs sample comprised of $12057$ asteroids.

In this study we used separate orbital sub-samples rather then the whole $\sim$$44000$ data set. We used the sub-samples because the orbits originated from different sources can have significantly different statistical properties. On Fig. \ref{fig:01} one can see that distribution of the orbital elements of the radar and video samples can differ significantly. Therefore mixing all orbital samples together (even obtained by the same technique), i.e. utilization of the integrated samples blurs their individual statistical properties.
\subsection{Synthetic orbital samples}
\label{uchacha}
To assess the probability of spurious pairing or more numerous clustering among the observed orbits one needs a sample of random orbits fulfilling two requirements:
\begin{enumerate}[i]
\item the sample should be free from clusters originated from the same parent body,
\item the orbital distributions of the random particles should be statistically compatible with the orbital distributions of the observed objects. 
\end{enumerate}
The first requirement is easy to fulfill and needs no explanation. However, in case of the second one, generation of the random orbits demands some care.

The orbits of the meteoroids and the NEAs we define by a set of Keplerian elements $1/a,q,e,\omega,\Omega$ and $i$. Among  $q, e, 1/a$ we have well known algebraic relation:
\begin{equation}
\label{row:02}
1/a=(1-e)/q.
\end{equation}
And additionally, due to the Earth crossing condition (observation selection effect) for vast majority of meteoroids we have also relationship among $q, e, \omega$, namely:
\begin{equation}
r={{q(1+e)}\over{1 \pm e\cos\omega}}
\label{row:03}
\end{equation}
where $r$ is the  heliocentric distance of the meteoroid at the moment of collision with Earth ($r \cong 1[AU]$); the
sign at the $e\cos \omega$ is positive for the negative geocentric ecliptic latitude of a meteor
radiant, \mbox{$\beta_R<0$.}

To find how these relations influence the assessment of the probability of spurious grouping, we generated artificial orbits by a few methods. As the core uniform random number generator, $U(0,1)$, we took the subroutine {\em ran2} from the Numerical Recipes v. 2.0  package \citep{Numreci} \\  
Each set of the orbital elements  $1/a,e,q,\omega,\Omega$ and $i$  was determined by one of the methods: \\
\noindent
\paragraph{Method A}
\vspace{-0.2cm}
\begin{enumerate}[i]
\item  Find $min$ and $max$ values of each orbital element in the observed sample, (the range values),
\item  using uniform number generator $U(min,max)$, generate each orbital element separately.
\end{enumerate}
\vspace{-0.1cm}
\paragraph{Method B}
\vspace{-0.2cm}
\begin{enumerate}[i]
\item  For given observed sample, obtain the histograms of the orbital elements,
\item  generate each element separately by  inversion of the cumulative probability distribution function (the CPD inversion method, see appendix A).
\end{enumerate}
\vspace{-0.1cm}
\paragraph{Method C}
\vspace{-0.2cm}
\begin{enumerate}[i]
\item   For given observed sample, obtain the histograms of the orbital elements: $q,e,i,\Omega,\omega$,
\item  generate each element separately by the CPD inversion method,
\item  using (\ref{row:02}) and generated values of $q, e$  calculate $(1/a)_c$,
\item  if $(1/a)_c > (1/a)_{max})$  repeat steps (ii), (iii), (iv).
\end{enumerate}
\vspace{-0.1cm}
\paragraph{Method D}
\vspace{-0.2cm}
\begin{enumerate}[i]
\item   Obtain the sample ranges of the elements $1/a, \omega, \Omega$ and 
        the histograms of elements $q,e,i$,
\item   generate $q, e, i$ separately by the CPD inversion method,
\item  using (\ref{row:02}) and generated values of $q, e$ calculate $1/a_c$,
\item  if $1/a_c > 1/a_{max}$; 
       repeat steps (ii), (iii), (iv),
\item  using corresponding ranges, generate $\omega, \Omega$ employing the uniform distribution $U(min,max)$.
\end{enumerate}
\vspace{-0.1cm}
\paragraph{Method E}
\vspace{-0.2cm}
\begin{enumerate}[i]
\item   For given observed sample obtain the histograms of the orbital elements and additionally 
        find the fraction of the orbits $\beta_{fr}$ for which the geocentric ecliptic latitudes of  the meteor radiant $\beta > 0 $,
\item  generate $ e, \omega, \Omega, i$ separately by the CPD inversion method,
\item  using equation (\ref{row:03}) and obtained $e, \omega$ calculate $q_c$; choose the sign 
       at the $e\cos \omega$ by generating the number $b$ distributed uniformly, $U(0,1)$;  
       if $b> \beta_{fr}$ choose sign ``$-$'' otherwise choose sign ``$+$'',        
\item  using $q_c$ and generated earlier $e$ calculate $1/a_c$, if $1/a_c > 1/a_{max}$,  
       repeat steps (i),(ii), (iii), (iv).
\end{enumerate}
Surely, vast majority of the orbits obtained by method $A$ do not fulfill the Earth crossing condition, thus in this sense method A gives many unrealistic orbits. We used this method because we were curious to know how the assessment of the probability of spurious pairing among the meteoroid or the NEA orbits will differ for this method and for the more realistic ones. \\
Method B is more realistic one, however it doesn't count any correlation among the orbital elements. Method C is more realistic then method B because generated values of $e$ and $q$ have to represent the orbit with realistic (possible to be observed) value of the semi-major axis $a$. 

\begin{figure}
\centerline{
\includegraphics[width=0.23\textwidth]{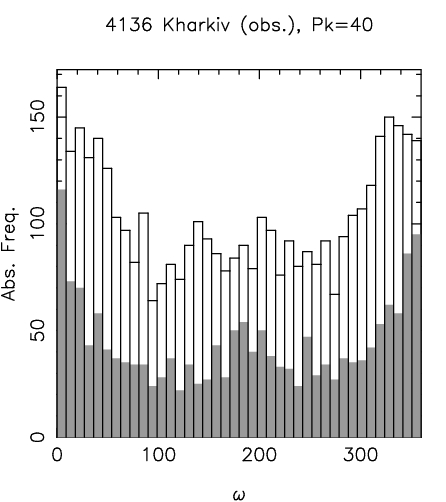}
\includegraphics[width=0.23\textwidth]{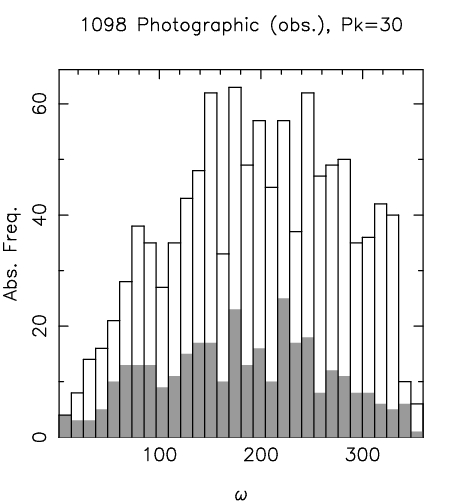}
}
\caption{Histograms of the argument of perihelion of sporadic meteoroid orbits. On the left hand panel distribution of 4136 orbits determined in Kharkov by radio technique. On the right hand panel distribution of $1098$  photographic orbits obtained by  Super-Schmidt and small camera equipment. Both distributions differ significantly from the uniform one. The gray beans correspond to the retrograde orbits.}
\label{fig:02}
\end{figure}

\noindent In the  case of method D we wanted to find how the assumption about the uniform distribution of the  $\omega$ and~$\Omega$ influences on  the probability of random grouping? 
 In case of the meteoroids, due to fast randomization of the $\omega$ and~$\Omega$ some authors drawn the  random values of these elements assuming uniform distributions of the observed values. Sometimes such assumptions is justified, however not always. On Fig. \ref{fig:02} we have shown  distributions of the observed values for the argument of perihelion  $\omega$ for the radar and photographic orbital samples. As one can see they significantly differ from the uniform distributions. 

\begin{figure}[t!]
\centerline{
\hbox{
\vbox{
\hbox{
\includegraphics[width=0.23\textwidth]{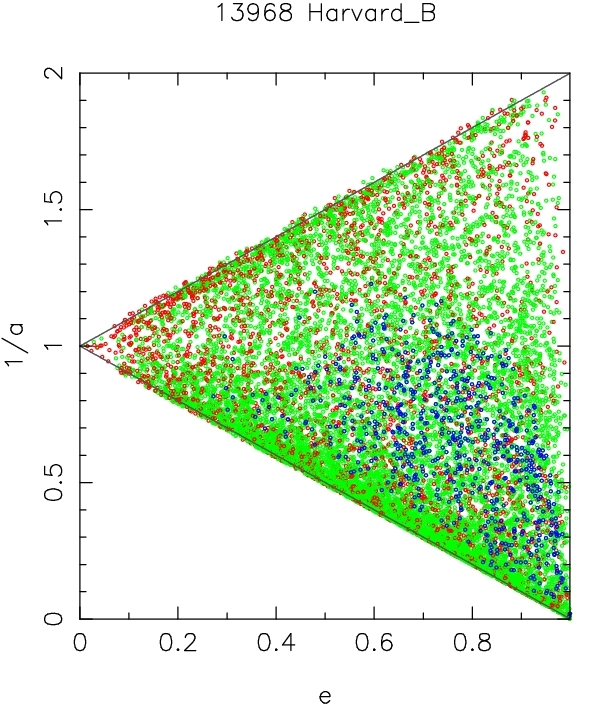}
\includegraphics[width=0.23\textwidth]{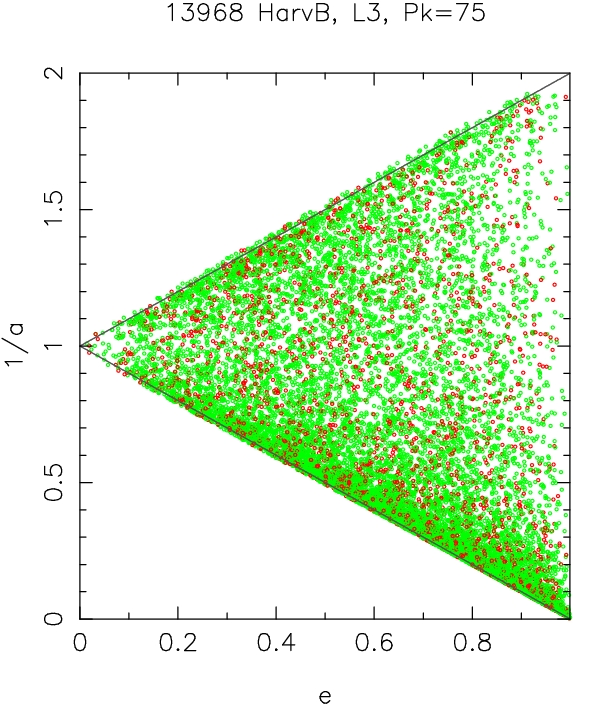} 
}
\vbox{
\hbox{
\includegraphics[width=0.23\textwidth]{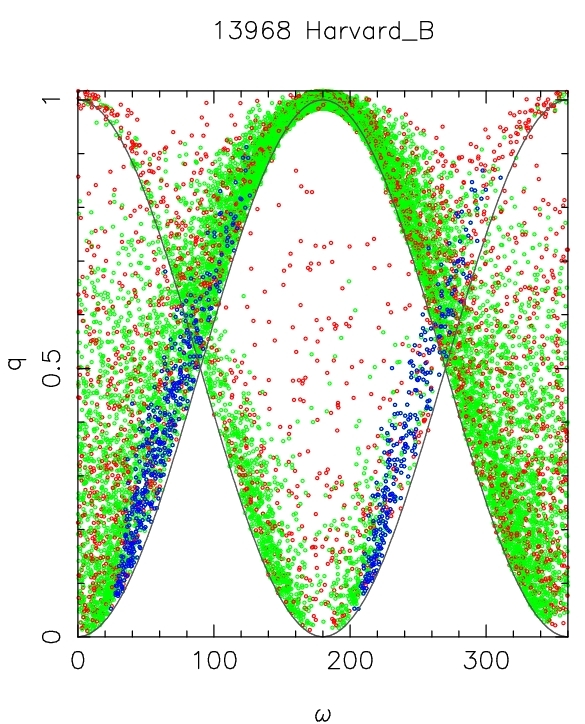}
\includegraphics[width=0.23\textwidth]{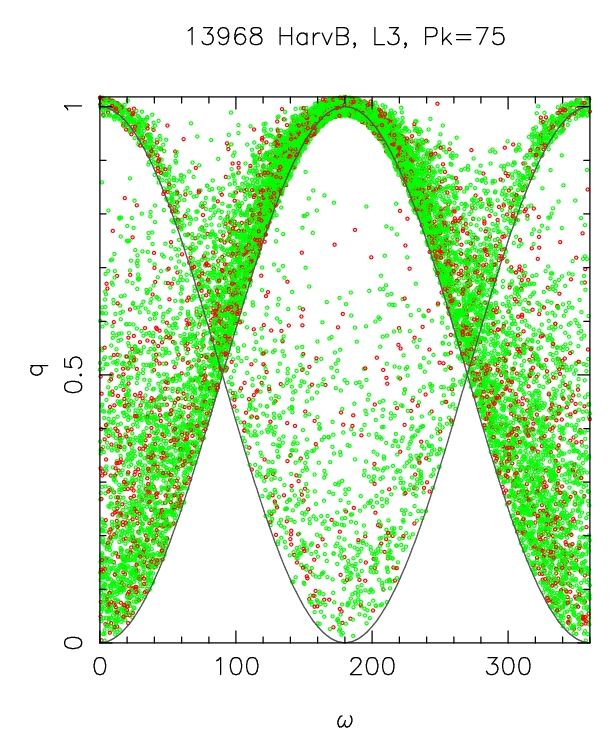}
}}
}}
}
\caption{Illustration of the quality of the artificial orbits  E-generator. On the left hand panels the plots relates to the observed Harvard synoptic year orbital sample. On the right hand panels the plots relates to the generated orbits. On the planes $(e-1/a)$ and  $(\omega - q)$ we can see two regions forbidden for meteors to be observed from Earth. The red color marks the orbits for which $i > 90\deg$. The blue points on the left marks daytime meteors. In case of the synthetic sample the daytime option was not implemented.}
\label{fig:03}
\end{figure}

From all methods defined above, method E gives the most realistic meteoroid orbits. The artificial orbital sample obtained by this approach is statistically very similar to the observed sample.
Good quality of this method is illustrated on Fig. \ref{fig:03} --- the distributions of  the artificial orbits resemble the observed ones.

In our study the orbital similarity among two orbits was calculated using five orbital elements $e,q,\omega,\Omega,i$.
\section{The D-functions of the orbital similarity}
As a quantitative measure of the similarity between two orbits we have used three distance functions: the $D_{SH}$ function introduced by \citet{SouthworthHawkins}, the $D_D$ function proposed by \citet{Drummond} and their hybrid $D_H$ by \citet{Jopek1993}.  Using  the unit vectors ${\bf \hat h}$ of the angular (orbital) momentum  and the Laplace unit vector  ${\bf \hat e}$, see e.g. \cite{Breiter05, Jopek2008}:
\begin{equation}
\mathbf{\hat h} =
\left(
\begin{array}{c}
  \sin i \sin\Omega \\
 - \sin i \cos\Omega \\
\cos i 
\end{array}
 \right),
\end{equation}
\begin{equation}
\mathbf{\hat e} =
\left(
\begin{array}{c}
\cos\omega\cos\Omega-\cos i \sin\omega\sin\Omega\\ 
\cos\omega\sin\Omega+\cos i \sin\omega\cos\Omega\\ 
\sin i \sin \omega
\end{array}
 \right),
\end{equation}
the distance $D_{SH}$ \citep{SouthworthHawkins} between two orbits $A$ and $B$ is given by:
\[
D^{2}_{SH} = \left(e_{B}-e_{A}\right)^{2}+\left(q_{B}-q_{A}\right)^{2}+
\]
\begin{equation}
\hspace{0.70cm}
+\left(2\sin\frac{I_{AB}}{2}\right)^{2}+
\left(\frac{e_{A}+e_{B}}{2}\right)^{2}\left(2\sin\frac{\pi_{AB}}{2}\right)^{2},
\label{row:dsh}
\end{equation}
where
\begin{equation}
\begin{array}{ll}
 I_{AB}  &= \arccos (\bf{\hat{h}}_A \cdot \bf{\hat{h}}_B) \\
\pi_{AB} &= \arccos(\bf{\hat{N}} \cdot \bf{\hat{e}}_A) - \arccos(\bf{\hat{N}} \cdot \bf{\hat{e}}_B)\\
\bf{N}   &= \bf{\hat{h}}_A \times \bf{\hat{h}}_B.         
\end{array}
\end{equation}
Vector ${\bf N} $ points to the common node of two orbits.
\\
\noindent
The distance function $D_D$ \citep{Drummond} is defined as:
\[
D_D^2 = \left( \frac{e_B-e_A}{e_B+e_A} \right)^2 + \left( \frac{q_B-q_A}{q_B+q_A} \right)^2
\] 
\begin{equation}
+ \left( \frac{I_{AB}}{180^\circ } \right)^2+ \left( \frac{e_B+e_A}{2} \right)^2 \left( \frac{\theta_{AB}}{180^\circ} \right)^2,
\label{row:dr}
\end{equation}
where
\begin{equation}
\begin{array}{ll}
I_{AB}  &= \arccos (\bf{\hat{h}_A} \cdot \bf{\hat{h}_B})\\
\theta_{AB} &=\arccos (\bf{\hat{e}_A} \cdot \bf{\hat{e}_B})
\end{array}
\end{equation}
The hybrid $D_H$ \citep{Jopek1993} is given by:
\[
D_{H}^2 = \left(e_B-e_A \right)^2 + \left( \frac{q_B-q_A}{q_B+q_A} \right)^2 + 
\]
\begin{equation}
\left( 2\sin \frac{I_{AB}}{2} \right)^2+ \left( \frac{e_B+e_A}{2} \right)^2 \left( 2 \sin \frac{\pi_{AB}}{2} \right)^2
\label{row:dh}
\end{equation}
\section{Assessments of the probability: two methods}
In practice, two orbits $(A,B)$ are defined as similar if their $D(A,B)$ value does not exceed certain threshold value $D_C$. Fixing the threshold $D_C$, for any given set of $O$ orbits drawn randomly, one can ask --- what is the probability  $P_2$ of finding among  $O$ orbits at least one pair $(A,B)$ such that  $D(A,B)<D_C$? To assess  $P_2$ (see the idea proposed in \citet{Jopek1997}) it is sufficient to generate $n$ sets of $O$ orbits and find a number of sets $k\leq n$ in which at least one similar pair $(A,B)$  was identified. Estimated probability is given as a relative frequency:
\begin{equation} 
P_2(D(A,B)< D_C)=k/n
\label{lulu}
\end{equation}
\cite{PaulsGladman} proposed interesting approach based on the geometric probability distribution i.e. the probability distribution of the number of $X$ Bernoulli trials needed to get one success. If the probability of success at first trial is $p$, then probability that the first success occurs at $k$-th trial is given by:
\[
P(X=k)=(1-p)^{k-1}p
\]
The expected value of a geometrically distributed random variable $X$ equals:
\begin{equation}
E(X)=1/p
\label{ro:ber}
\end{equation}
Thus, defining a success as supervention (among $O$ orbits) of a pair $(A,B)$ for which $D(A,B)<D_C$, and finding the average number of trials required for a success, probability $P_2(D(A,B)<D_C)$ of a random similarity between $A$ and $B$ orbits can be estimated by:
 \begin{equation}
P_2(D(A,B)< D_C)= p= 1/E(X)
\label{ro:praw}
\end{equation}
Both above approaches  can easily be extended for groups of 3, 4, 5  or more members. 
\subsection{Probability assessment:  numerical experiments}
In this work, we made use of equations (\ref{lulu}, \ref{ro:praw}). For a given number of $O$ orbits we have drawn equally numerous set of $O$ synthetic orbits and among them we searched for a first pair $(A,B)$ for which $D(A,B)<D_C$. Drawings of $O$ orbits and searching for a similar pair was repeated $k$ times until the first such pair occurred, and the number $k$ was recorded. Such process was repeated $n=500$ times and then the expected value $E(k)$ was calculated. Finally, using equation (\ref{ro:praw}), the probability of  random paring between two orbits was estimated. This estimate corresponds to the assumed threshold $D_C$ and the size $O$ of the orbital sample.
 
Similarly, when equation (\ref{lulu}) was applied, for given $D_C$, $n=5000$ sets of $O$ synthetic orbits were genera\-ted and searched for at least one pair $(A,B)$ for which $D(A,B)<D_C$. We counted the number $k$ of sets of $O$ orbits among which a pair was found. Afterwards probability $P_2(D(A,B)<D_C)$ was calculated by equation (\ref{lulu}).

In both approaches we tested how obtained probability depends: on the choice of the core uniform U(0,1) random number generator (primacy was given to  $ran2$ routine taken from \cite{Numreci}); on the initializing  seed of the U(0,1) generator; on the number $n$ of the repetition of the Bernoulli trials or the  number of $n$ generated sets of $O$ orbits; on amount of beans in the histograms of the observed orbital elements.       
We have found that in each approach the joint influence of the above factors was similar and the accuracy of the determined probabilities was better than $0.005-0.01$ for $P_2 < 0.2$. For the larger probabilities the accuracy was equal $0.01-0.02$.
\subsection{Comparison of two methods}
We made comparison of the results obtained by both approaches. The histograms of bolides orbits (see Table \ref{ta:01}) were used for generating (by method A) the synthetic orbits.  For a few values of thresholds $D_C$ and the $D_{SH}$ distance function several probabilities $P_2(D(A,B)<D_C)$ of random pairing were found among $200$ and $800$ synthetic bolides orbits.  Obtained results are given in Table~\ref{ta:02}. We found that probabilities assessed by Bernoulli trials method (BT) and by the relative frequency approach (RF) are equivalent. Results given by these methods differ on the level of the accuracy of each method. However as to the computing  time required for calculations the RF method proved to be superior, on average it was by 1-2 orders of magnitude faster. For this reason, in the further study we have used the RF method only.
\begin{table}
\small
\caption[] {Probabilities of a random pairing among $200$ and $800$ synthetic bolides orbits assessed by Bernoulli trials  method (BT) and by the relative frequency approach (RF). Three orbital similarity thresholds $D_C$  were assumed and two orbital samples were searched for similar pairs.  In column RF-BT the differences between the probabilities obtained by RT and BT methods are given. The synthetic orbits were generated by method A using histograms ($20$ bins) of the observed bolide's orbital elements. The orbital similarity was determined by $D_{SH}$ distance function.}
\label{ta:02}
\begin{center}
\begin{tabular}{|c|c|c|c|c|}
\hline
\multicolumn{1}{|c|}{ } & \multicolumn{2}{c|}{200 orbits} & \multicolumn{2}{c|}{800 orbits}\\
\hline
\multicolumn{1}{|c|}{$D_C$} & \multicolumn{1}{c|}{RF} & \multicolumn{1}{c|}{RF-BT} & \multicolumn{1}{c|}{RF} & \multicolumn{1}{c|}{RF-BT}\\
\hline
0.09  & 0.2482  &  0.0015 & 0.9904  & -0.0017 \\
0.05  & 0.0222  & -0.001  & 0.2894  & -0.0098 \\
0.02  & 0.0002  & -0.0003 & 0.0034  &  0.0009  \\
\hline
\end {tabular}
\end {center}
\normalsize
\end {table}
\section{The results: probability of a random pairing}
The probability of a random pairing, among other things, depends on the distance function $D(A,B)$, the  orbital similarity threshold value $D_C$ as well as on the size and origin of the orbital sample. In our study the origin of the orbital sample implies the orbital distribution of the observed sample and the method by which the orbits were generated (see section \ref{uchacha}).

For a given distance function $D(A,B)$ and the orbital sample our main objective was to find the threshold values $D_C$ for which the probabilities $P_2$ of a random similarity among two orbits  ($D(A,B) < D_C$) were sufficiently small and were equal to $0.01$--$0.05$.
\begin{figure}[t!]
\centerline{
\vbox{
\includegraphics[width=0.4\textwidth]{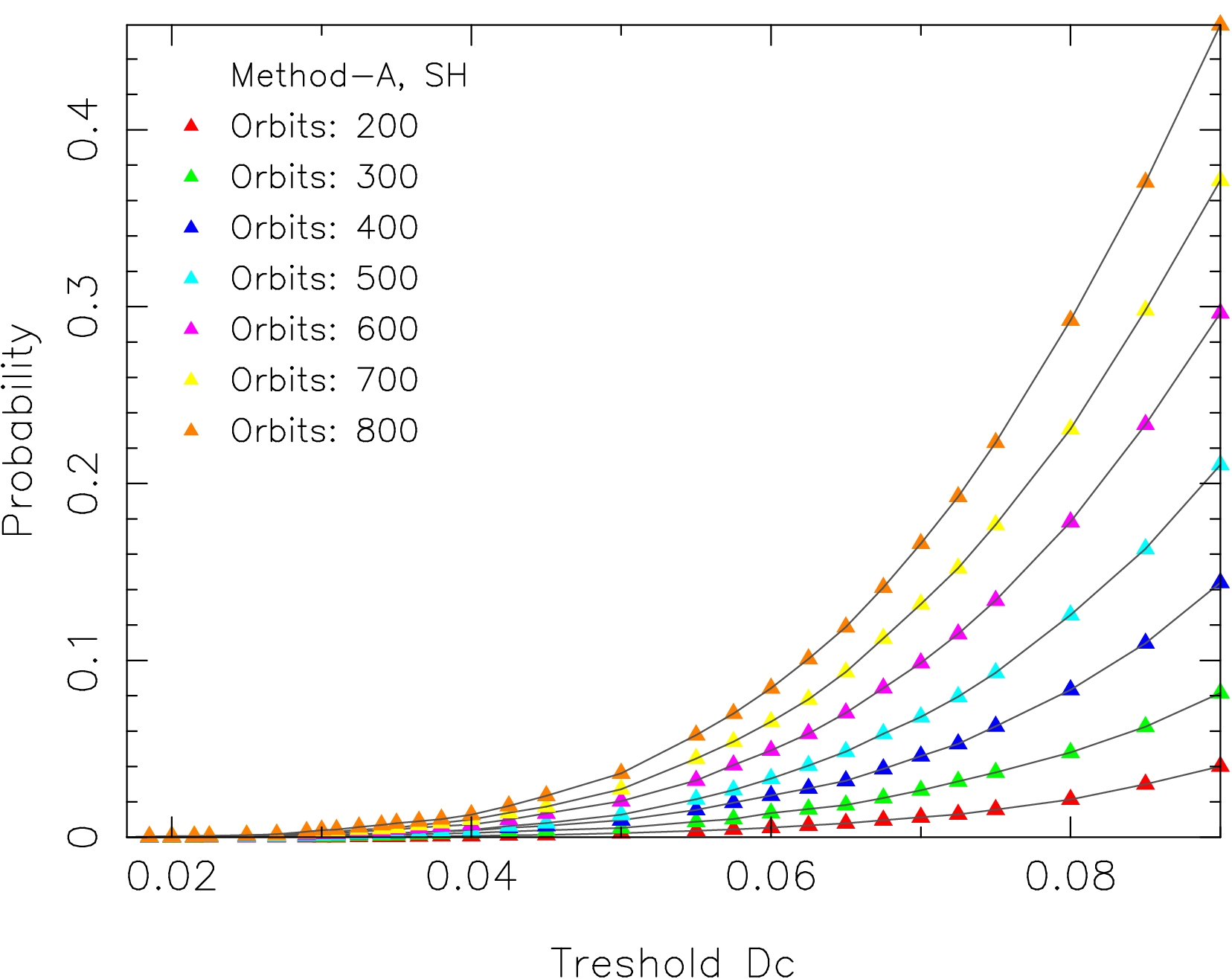}
\vspace{0.2cm}
\includegraphics[width=0.4\textwidth]{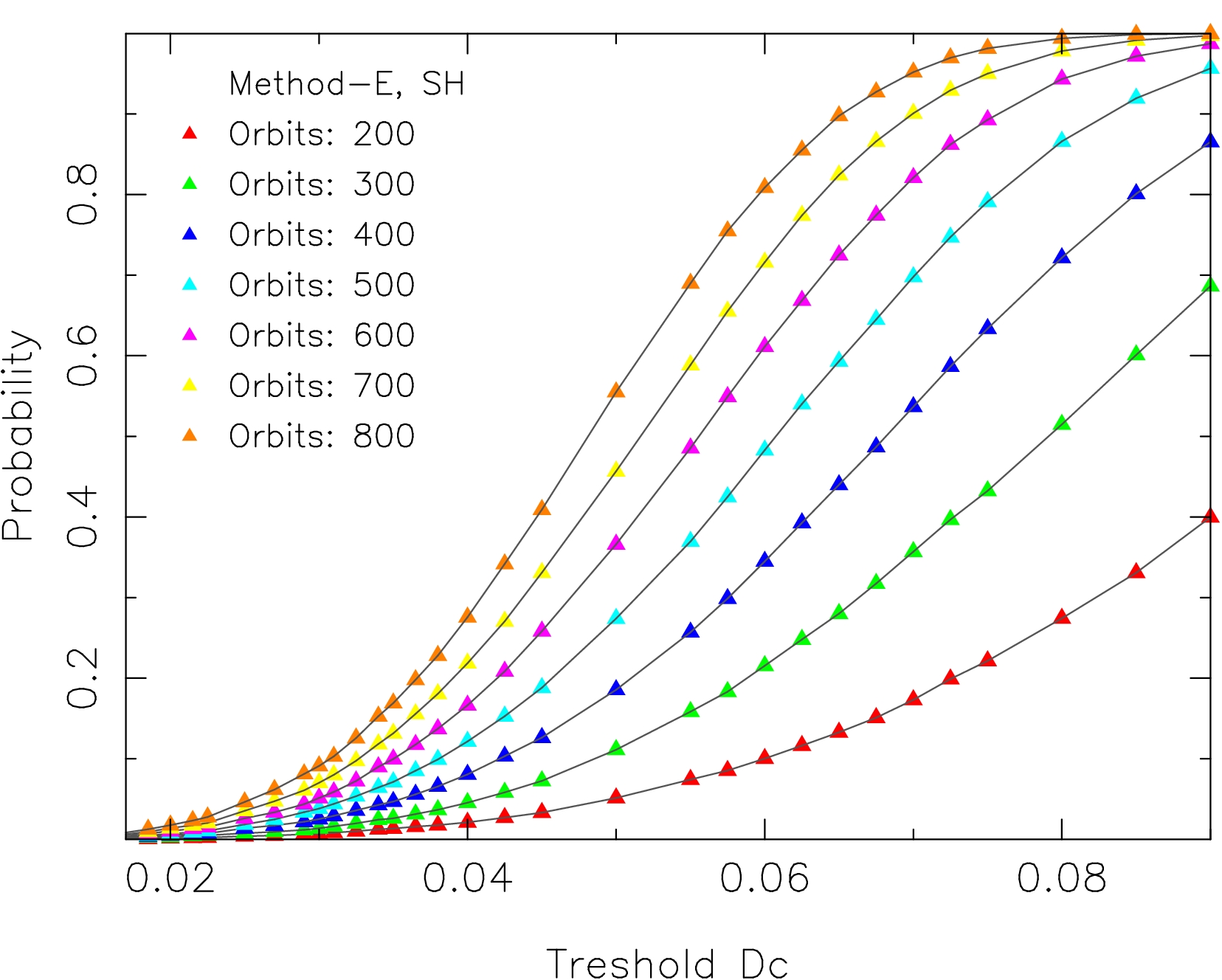}
}
}
\caption{Probabilities of a random pairing for different orbital similarity thresholds $D_C$ and different orbital sample sizes obtained with the  $D_{SH}$ distance function (equation (\ref{row:dsh})).
The plots present the results obtained for synthetic bolide's orbits generated by the method A (the top panel) and the method E (the bottom panel) (see section \ref{uchacha}). The curves show similar course, but for a fixed threshold e.g. $D_C=0.06$ the values of the probabilities visible on both panels differ significantly. 
 }
\label{P2lolo}
\end{figure}
Firstly, for a given $D$-function, we found how the probability $P_2$ changes with $D_C$  and the orbital sample size and its origin. On Figure~\ref{P2lolo} we illustrate such relationship and one can see that the curves on these graphs differ significantly. E.g. in case of the method A, for $800$ bolides orbits and $D_C=0.06$, probability of pairing $P_2\approx0.1$. In case of method E this probability $P_2\approx0.8$. It should be so, because having the same amount of orbits distributed uniformly in a greater volume of the orbital elements space (free from forbidden regions, method A), the density of  points (orbits) is smaller and consequently, the average distance between two points is greater. As result, in such volume we have lesser chance to find a close orbital pair. When method E is used, some parts of the orbital element's volume are forbidden due to the Earth crossing condition for the observed meteoroid's orbits (see Fig.~\ref{fig:03}), it results in higher density of the points and the average distance between two points is smaller.  
\begin{figure}[t!]
\centerline{
\includegraphics[width=0.4\textwidth]{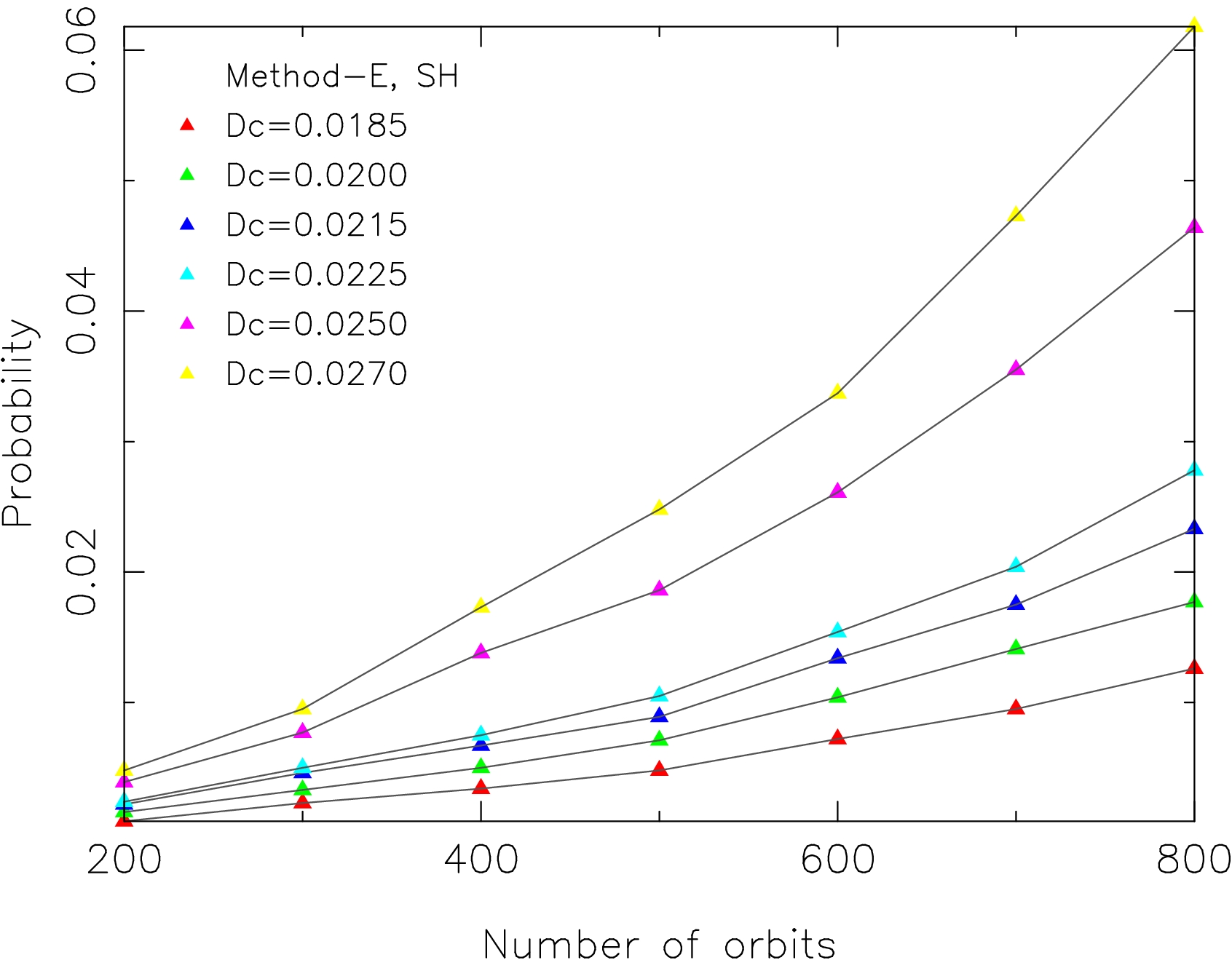}
}
\caption{Probabilities $P_2$  of random pairing versus the size of the orbital sample for a few fixed thresholds $D_C$. As can be seen this dependence is not linear. The curves correspond to the bolide's orbits, $D_{SH}$ function and method E. }
\label{P2vN}
\end{figure}
\begin{figure}[t!]
\centerline{
\includegraphics[width=0.4\textwidth]{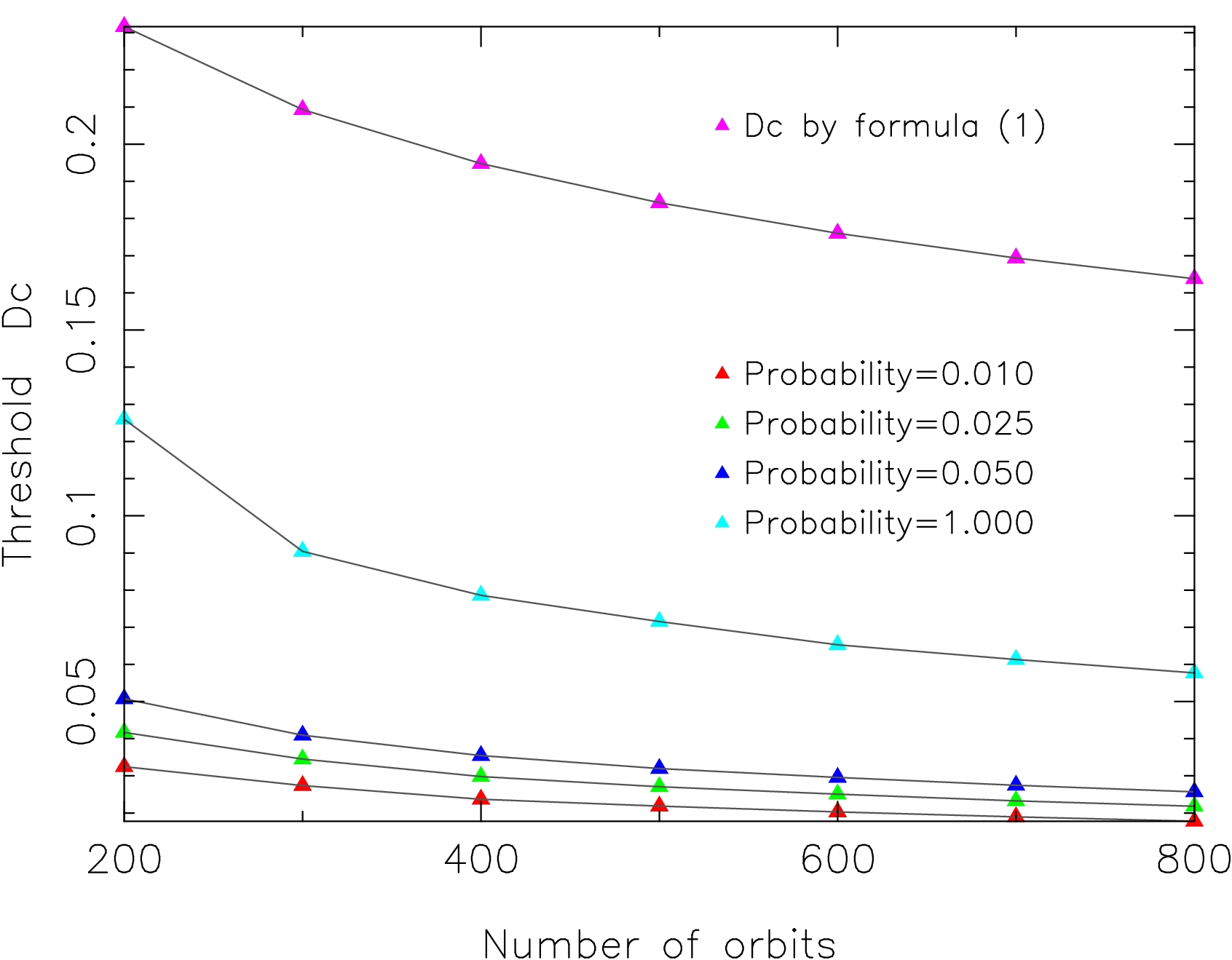}
}
\caption{Results for the bolide's orbits, $D_{SH}$ function and method E. The orbital similarity thresholds $D_C$ versus size of the orbital sample for a few values of probabilities of a random pairing. 
All values of the thresholds  $D_C$ above the curve market with cyan triangles correspond to the probability of random pairing equals 1.0. As one can see, for bolides orbits this conclusion involves all $D_C$ values calculated by formula~(\ref{row:01}). 
Some data plotted on this graph are given in Table~\ref{Bolid_Dc}.
 }
\label{P2DcvN}
\end{figure}
\begin{table}[t!]
\small
\caption[] {$D_{SH}$ function; method E. The orbital similarity thresholds $D_C$ for three values of the probability of a random pairing $P_2$ among the synthetic bolide's orbits. The orbital samples were varied by the size $N$. (See Figure \ref{P2DcvN}). The last columns contain thresholds $D_{(1)}$, $D_{(\ref{row:DC_SH_bol})}$  calculated by formulae (\ref{row:01}) and (\ref{row:DC_SH_bol}), respectively. 
}
\label{Bolid_Dc}
\footnotesize
\begin{center}
\begin{tabular}{|c|c|c|c|c|c|  }
\hline
\multicolumn{1}{|c|}{$P_2=$ } & \multicolumn{1}{c|}{$0.01$} & \multicolumn{1}{c|}{$0.025$} & \multicolumn{1}{c|}{$0.05$}  & \multicolumn{1}{c|}{Form. (\ref{row:01})}  & \multicolumn{1}{c|}{Form. (\ref{row:DC_SH_bol})}\\
\hline
\multicolumn{1}{|c|}{$N$ } &\multicolumn{1}{|c|}{$D_{C1}$} & \multicolumn{1}{c|}{$D_{C2}$} & \multicolumn{1}{c|}{$D_{C3}$}  & \multicolumn{1}{c|}{$D_{(1)}$} &  \multicolumn{1}{c|}{$D_{(\ref{row:DC_SH_bol})}$}                                         \\
\hline
 200 &   0.0325 &  0.0417  &  0.0507 & 0.2317  &0.0324 \\ 
 300 &   0.0274 &  0.0345  &  0.0409 & 0.2093  &0.0272 \\ 
 400 &   0.0237 &  0.0298  &  0.0354 & 0.1948  &0.0241  \\ 
 500 &   0.0218 &  0.0271  &  0.0320 & 0.1842  &0.0218 \\ 
 600 &   0.0203 &  0.0251  &  0.0296 & 0.1760  &0.0202 \\ 
 700 &   0.0190 &  0.0233  &  0.0274 & 0.1694  &0.0189 \\ 
 800 &   0.0178 &  0.0218  &  0.0257 & 0.1638  &0.0178 \\ 
\hline
\end {tabular}
\end {center}
\normalsize
\end {table}
\newline
On Figure \ref{P2vN} we illustrate how the $P_2$ probabilities change with increase of the size of the orbital sample; the curves are plotted for several constant thresholds $D_C$. Keeping fixed $D_C$, the increase of the sample size (i.e. increase of the orbital density in the fixed volume) leads to increase of the value of the probability of random pairing. The growth rate is small for small $D_C$. For large  $D_C$ the probability  $P_2$ increases much faster with the sample size, even by factor $2$-$3$.

Figure \ref{P2DcvN} presents results of the Least Square Fitting  (LSF) applied to the data illustrated on the bottom panel of Figure \ref{P2lolo}. The polynomials of degrees $4$-$5$-th were fitted to the data points for  each curve separately. For calculations we used the subroutine $lfit$ from \cite{Numreci}. Because we were interested in small values of $P_2$, only the first part of curves $P_2(D_C)$ were fitted using adequately selected pairs ($D_C, P_2$). In the next step the polynomials were used to find $D_C$ values corresponding to  the fixed values of $P_2=\{ 0.01, 0.025, 0.05\}$. Obtained results are given in Table \ref{Bolid_Dc}. 

The contents of Table \ref{Bolid_Dc} has some practical value. For example, in the range of the orbital sizes $200$--$800$ and for the probability  $P_2=0.01$ and $D_{SH}$ function it can be used for assessment of the proper threshold $D_C$. Of course, provided that the orbital sample has the orbital distributions similar to the bolide's sample used in this study.
\subsection{Influence of the method used for synthetic orbits generation}
 In the past the probability of a random grouping were estimated using different methods for generating the synthetic orbits. E.g. in \cite{Jopek1997, Jopek2003} method B was used for the meteoroid orbits, in \cite{Nesvorny2006} the authors used method A for the asteroids orbits. The plots on Figure \ref{P2lolo} illustrate that for given value $D_C$ and the fixed orbital sample size, the values of the probabilities of a random pairing clearly depend on the method used for generating the synthetic orbits. In this study have tested how the thresholds $D_C$ corresponding to small probability  $P_2=0.01$ depend on the method used for generating the synthetic orbits. The results are given in Table \ref{P2pormet}. This Table contains the values of thresholds $D_C$ for the bolide like orbits generated by the methods described in section \ref{uchacha}.
The results clearly show that obtained thresholds can differ significantly. In particular, method A gives results significantly different  (by factor 2) from these obtained with method E. Methods B,C,D give comparable results, the differences are on the level of the accuracy of our calculations. However these results are clearly different from the thresholds obtained with method E.
Hence, bringing back the remarks written at the end of section \ref{uchacha}, obtained result convinced us that for meteoroids,  to estimate the correct values of the orbital similarity thresholds one should use the method E. In case of the meteoroids orbits this method was used in our further studies solely. 
\begin{table}[t!]
\small
\caption[] {$D_{SH}$ function; bolides orbital sample. The thresholds $D_C$ for the fixed probability $P_2=0.01$ of a random pairing among the synthetic bolide's orbits. The orbital samples, varying sizes $N$, were generated by five methods A,B,C,D,E described in section \ref{uchacha}.}
\label{P2pormet}
\begin{center}
\begin{tabular}{|c|c|c|c|c|c|}
\hline
\multicolumn{1}{|c|}{ } & \multicolumn{5}{c|}{$D_C$ for $P_2=0.01$} \\
\hline
\multicolumn{1}{|c|}{$N$ } &\multicolumn{1}{|c|}{A} & \multicolumn{1}{c|}{B} & 
\multicolumn{1}{c|}{C} & \multicolumn{1}{c|}{D} & \multicolumn{1}{c|}{E}   \\
\hline
 200 &   0.0683 &  0.0422  &  0.0418  &  0.0418  &  0.0325   \\ 
 300 &   0.0567 &  0.0359  &  0.0348  &  0.0357  &  0.0274  \\ 
 400 &   0.0498 &  0.0323  &  0.0304  &  0.0324  &  0.0237 \\ 
 500 &   0.0473 &  0.0290  &  0.0282  &  0.0297  &  0.0218 \\ 
 600 &   0.0426 &  0.0262  &  0.0253  &  0.0279  &  0.0203 \\ 
 700 &   0.0398 &  0.0243  &  0.0243  &  0.0257  &  0.0190 \\ 
 800 &   0.0373 &  0.0237  &  0.0233  &  0.0248  &  0.0178 \\ 
\hline
\end {tabular}
\end {center}
\normalsize
\end {table}
\begin{figure}[b!]
\centerline{
\includegraphics[width=0.5\textwidth]{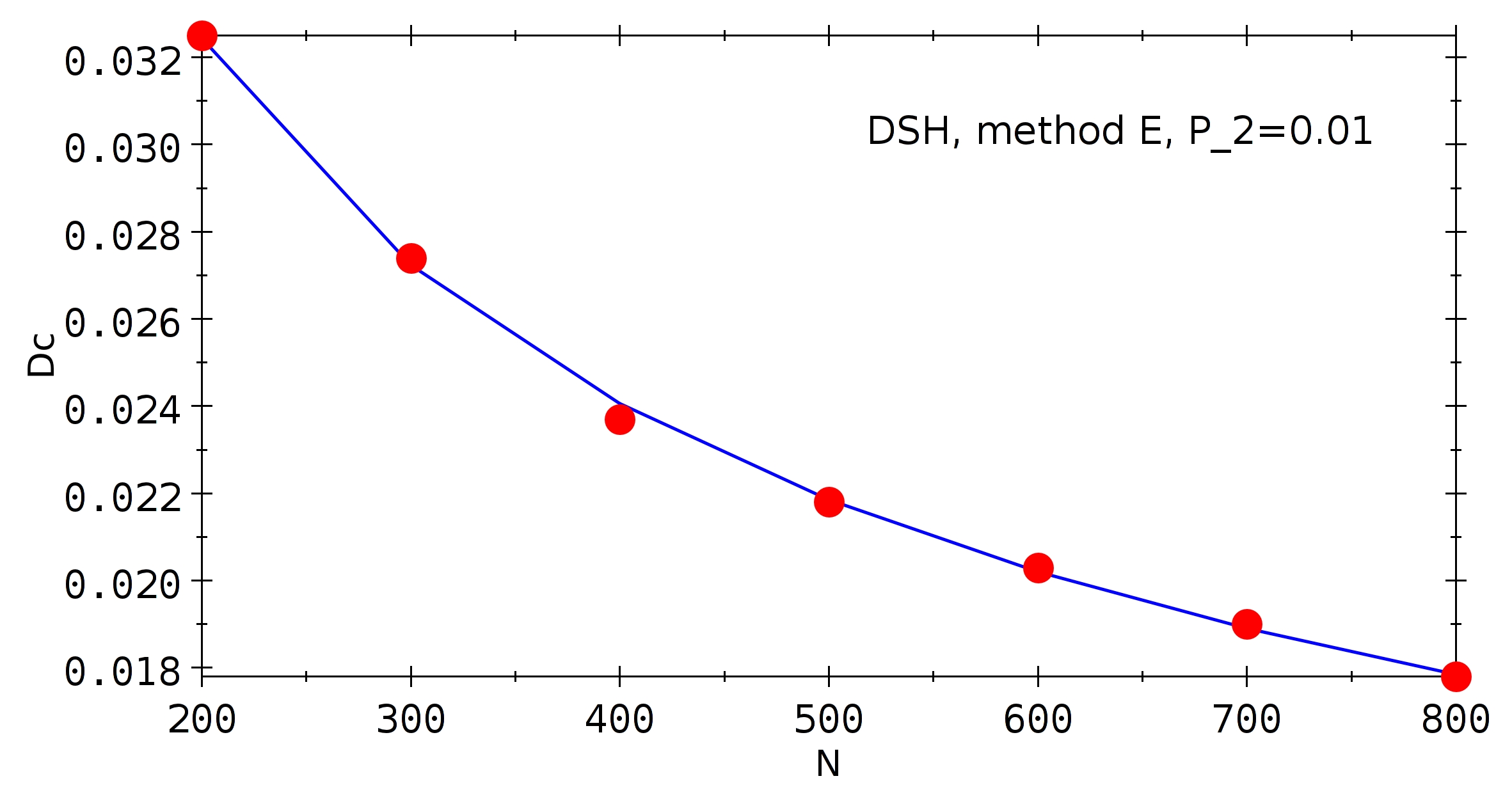}
}
\caption{Thresholds $D_{C}$ versus orbital sample size $N$ for the constant value of the probability $P_2=0.01$ of a random orbital similarity. The red points correspond to the $N$ and $D_{C1}$ values taken from Table \ref{Bolid_Dc}. The blue curve was obtained by the method of least squares. This dependence is given by formula (\ref{row:DC_SH_bol}).}
\label{fig:DC_N_SH}
\end{figure}
%
%
\subsection{Orbital pairing: bolide samples}
On Figure \ref{P2DcvN} and also in Table \ref{Bolid_Dc} we included the values of thresholds calculated by formula (\ref{row:01}). It turned out that these values are much to high in the context of a random pairing. On Figure \ref{P2DcvN} the curve corresponding to formula (\ref{row:01}) lies in the region in which probability $P_2=1$. That would imply the formula  (\ref{row:01}) should not be used for searching similar orbital pairs among the bolide samples. 

We propose its modification which gives the thresholds corresponding to the probabilities of coincidental similarity $P_2$ close to $0.01$. As one can see on Fig. \ref{P2DcvN} and Fig \ref{fig:DC_N_SH}, the dependence $D_{C}$ versus $N$ are not linear. However, using the logarithms of the data from first two columns of Table \ref{Bolid_Dc} by the LSF fitting we obtained modification of formula (\ref{row:01}), namely:
\begin{equation}
D_c=0.3186\cdot N^{-0.431}
\label{row:DC_SH_bol}
\end{equation}
For given $N$ formula (\ref{row:DC_SH_bol}) can be used for calculating the thresholds $D_{C}$ corresponding to $P_2 \approx 0.01$. Starting from $N=2$ it gives the value of threshold $D_C=0.2363$. For very big values of $N$ the thresholds asymptotically lead up to zero. 
The quality of this fitting is illustrated on Fig. \ref{fig:DC_N_SH} and in Table \ref{Bolid_Dc} where in the last column we placed the values of thresholds obtained by formula (\ref{row:DC_SH_bol}). The accordance between the thresholds given in columns $D_{C1}$ and $D_{(\ref{row:DC_SH_bol})}$ is very good.

Of course, formula (\ref{row:DC_SH_bol}) is valid only for searching pairs among  the bolide-type orbits by means of the $D_{SH}$ function. We believe that our formula can be used for the bolides samples of $200$--$1000$ or more orbits.
 
We repeated our approach with $D_D$ and $D_H$ functions defined by equations  (\ref{row:dr}) and (\ref{row:dh}). Obtained thresholds for the constants $P_2$ equal $0.01$ and $0.025$ are listed in Table~\ref{Bolid_DH_DR}. Analogously as for $D_{SH}$, also for $D_{H}$ and $D_D$ functions we found the formulae for calculation of the thresholds $D_c$ corresponding to the probability $P_2$ respectively: 
\begin{equation}
D_c=0.3143\cdot N^{-0.438}
\label{row:DC_HY_bol}
\end{equation}
\begin{equation}
D_c=0.1240\cdot N^{-0.423}
\label{row:DC_DR_bol}
\end{equation}
In case of the bolide samples the thresholds for $D_{SH}$ and $D_H$ functions differ a little. For $D_{SH}$ function, on average they are  $1.05$ times higher than for $D_H$. In case of the $D_D$ function the thresholds are $2.45$ times smaller than for the $D_{SH}$ function.  
\begin{table}[t!]
\small
\caption[] {$D_{H}$ and $D_D$ functions; method E. The orbital similarity thresholds $D_C$ for the probability of a random pairing $P_2$ among the synthetic bolide's orbits. 
}
\label{Bolid_DH_DR}
\footnotesize
\begin{center}
\begin{tabular}{|c|c|c||c|c|}
\hline
\multicolumn{1}{|c|}{ } &\multicolumn{2}{|c||}{$D_{H}$} & \multicolumn{2}{c|}{$D_D$} \\
\hline
\multicolumn{1}{|c|}{$P_2=$ } & \multicolumn{1}{c|}{$0.01$} & \multicolumn{1}{c||}{$0.025$} & \multicolumn{1}{c|}{$0.01$}  & \multicolumn{1}{c|}{$0.025$} \\
\hline
\multicolumn{1}{|c|}{$N$ } &\multicolumn{1}{|c|}{$D_{C1}$} & \multicolumn{1}{c||}{$D_{C2}$} & \multicolumn{1}{c|}{$D_{C1}$}  & \multicolumn{1}{c|}{$D_{C2}$}\\
\hline
200 &   0.03123 &  0.04036  & 0.01333 &  0.01707   \\ 
300 &   0.02599 &  0.03239  & 0.01099 &  0.01374   \\ 
400 &   0.02218 &  0.02781  & 0.00978 &  0.01204   \\ 
500 &   0.02075 &  0.02503  &  0.00874 &  0.01073  \\ 
600 &   0.01944 &  0.02333  & 0.00837 &  0.01021  \\ 
700 &   0.01756 &  0.02137  &  0.00761 &  0.00934  \\ 
800 &   0.01712 &  0.02050  & 0.00748 &  0.00910  \\ 
\hline
\end {tabular}
\end {center}
\normalsize
\end {table}
\subsection{Orbital pairing for NEAs, bolide, radio and video small samples}
\label{smallsample}
\begin{figure}
\centerline{
\includegraphics[width=0.33\textwidth]{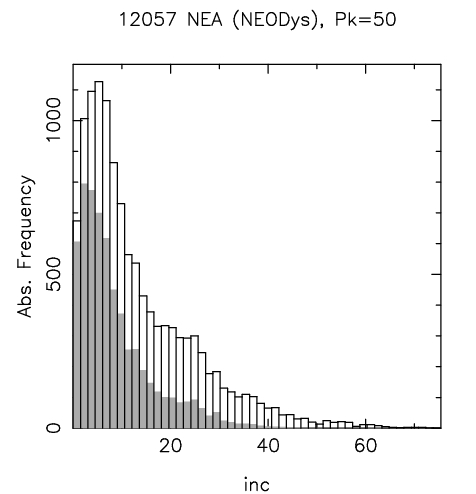}
}
\caption{Histogram of the orbital inclinations of the NEAs.
In this study we have used $12057$ orbits with $i<75.41$ degree, only.
The shadowed beans correspond to the NEAs for which the absolute magnitudes $H$ were greater than the sample median value, i.e. $H=21.96$.
}
\label{fig:inc}
\end{figure}

In this section we compare the values of thresholds obtained for different orbital samples, namely for the NEAs, bolides, radar Harvard B meteors and video SonotaCo 2009 meteoroid sample (see Table \ref{ta:01}). We made similar study as in the previous section, the results are given in Table~\ref{compsmall}. 
\begin{table*}[t!]
\small
\caption[] {$D_{SH}$ function; method E. The orbital similarity thresholds $D_C$ for the probability of a random pairing $P_2$ for the small samples of the synthetic NEAs, bolides, radar (Harvard B sample) and video (SonotaCo 2009) orbits. For the reader convenience we placed here some results for the bolides from Table \ref{Bolid_Dc}.  
}
\label{compsmall}
\footnotesize
\begin{center}
\begin{tabular}{|c|c|c|c|c|c|c|c|c|}
\hline
\multicolumn{1}{|c|}{ } &\multicolumn{2}{|c|}{NEAs} &  \multicolumn{2}{c|}{Bolides}& \multicolumn{2}{c|}{Harvard B}  & \multicolumn{2}{c|}{SonotaCo 2009}\\
\hline
\multicolumn{1}{|c|}{$P_2=$ } & \multicolumn{1}{c|}{$0.01$} & \multicolumn{1}{c|}{$0.025$} & 
\multicolumn{1}{c|}{$0.01$} & \multicolumn{1}{c|}{$0.025$} & 
\multicolumn{1}{c|}{$0.01$}  & \multicolumn{1}{c|}{$0.025$} & \multicolumn{1}{c|}{$0.01$}  & \multicolumn{1}{c|}{$0.025$} \\
\hline
\multicolumn{1}{|c|}{$N$ } & \multicolumn{1}{|c|}{$D_{C1}$} & \multicolumn{1}{c|}{$D_{C2}$} & 
 \multicolumn{1}{|c|}{$D_{C1}$} & \multicolumn{1}{c|}{$D_{C2}$} &
\multicolumn{1}{c|}{$D_{C1}$}  & \multicolumn{1}{c|}{$D_{C2}$} & \multicolumn{1}{c|}{$D_{C1}$}  & \multicolumn{1}{c|}{$D_{C2}$}\\
\hline
200 &  0.0293 &  0.0349 &  0.0325 & 0.0417 & 0.0381 &  0.0487  & 0.0416 &  0.0522   \\ 
300 &  0.0237 &  0.0297 &  0.0274 & 0.0345 & 0.0321 &  0.0393  & 0.0332 &  0.0415  \\ 
400 &  0.0217 &  0.0266 &  0.0237 & 0.0298 & 0.0275 &  0.0339  & 0.0301 &  0.0370  \\ 
500 &  0.0196 &  0.0247 &  0.0218 & 0.0271 & 0.0241 &  0.0309  & 0.0274 &  0.0333  \\ 
600 &  0.0189 &  0.0226 &  0.0203 & 0.0251 & 0.0227 &  0.0281  & 0.0248 &  0.0301 \\ 
700 &  0.0172 &  0.0210 &  0.0190 & 0.0233 & 0.0211 &  0.0261  & 0.0234 &  0.0286  \\ 
800 &  0.0163 &  0.0199 &  0.0178 & 0.0218 & 0.0192 &  0.0243  & 0.0218 &  0.0264 \\ 
\hline
\end {tabular}
\end {center}
\normalsize
\end {table*}
In Table \ref{compsmall} the orbital similarity thresholds corresponding to fixed $N$ and $P_2$, are the least for the synthetic NEAs orbits. In the ascending order, the corresponding thresholds are grater for bolides, Harvard B radar and SonotaCo 2009 video synthetic orbits. The least values for the NEAs arose from the smallest volume in the orbital elements space  occupied by this population. Despite that the NEAs orbit used in this study can have greater perihelion distance than the orbits of the meteoroids, their inclinations occupy much smaller interval. Their inclinations are never greater than $75.41$ degree, see Fig. \ref{fig:inc}. And this factor is the most significant at this point.

On average, the NEAs thresholds for $P_2=0.01$  proved to be $1.11, 1.25, 1.34$ times smaller then for the bolides, Harvard B radar, SonotaCo 2009 orbital samples, respectively. 
The thresholds for SonotaCo 2009 sample are (on average) $1.07$ times greater than corresponding thresholds for Harvard radar orbits. 

However the results presented here refer to small  sets of $200$-$800$ orbits. Because the databases containing the NEAs, radar and video orbits are much more numerous,  we have not found the formulae similar to those given by equations (\ref{row:DC_SH_bol}), (\ref{row:DC_HY_bol}), (\ref{row:DC_DR_bol}) which are valid for the samples of $200$-$800$ orbits only.
Corresponding  formulae we derived in the next section using the results for much bigger orbital samples.  Due to the same reasons we do not include here the results for NEAs, radar and video data obtained for $D_{H}$ and $D_{D}$ functions.    
\subsection{Orbital pairing for NEAs, radar and video large  samples}
In this section we present tables and practical formulae which can be used for finding the orbital similarity thresholds between two objects falling to the samples containing $1000$-$16000$ orbits. We present the results for different populations of the small bodies and three D-functions: $D_{SH}$, $D_H$ and $D_{D}$. In these calculations all synthetic meteoroid orbits were generated by the method $E$; in case of the NEAs population,  as the most appropriate,  the method $D$ was applied (see section \ref{uchacha}). Obtained results are given in Table \ref{complarge}. 
General properties of the thresholds given in Table \ref{complarge} are the same as in case of Table \ref{compsmall}; for fixed: D-function, the sample size $N$ and the probability $P_2$ -- the least thresholds we have for the NEAs orbits. For $D_{SH}$ function the thresholds $D_{C1}$ for radar (Harvard B) and video (SonotaCo2009)  data are (on average) 1.08 and 1.15 times greater than for the NEAs orbit. Average ratio between the thresholds for SonotaCo 2009 orbits and the Harvard B equals 1.06. These values are a bit smaller than corresponding ratios obtained in the previous section for the small orbital samples.
\begin{table*}[t!]
\small
\caption[] {The orbital similarity thresholds $D_C$ for the probability of a random pairing $P_2$ for the synthetic NEAs, radar (Harvard B sample) and video (SonotaCo 2009) orbital samples. The results were obtained for $D_{SH}$, $D_H$, $D_D$ functions and big orbital samples. 
}
\label{complarge}
\footnotesize
\begin{center}
\begin{tabular}{|c|c|c|c|c|c|c|}
\hline
\multicolumn{1}{|c|}{ } &\multicolumn{2}{|c|}{NEAs} & \multicolumn{2}{c|}{Harvard B}  & \multicolumn{2}{c|}{SonotaCo 2009}\\
\hline
\multicolumn{1}{|c|}{$P_2=$ } & \multicolumn{1}{c|}{$0.01$} & \multicolumn{1}{c|}{$0.025$} &  
\multicolumn{1}{c|}{$0.01$}  & \multicolumn{1}{c|}{$0.025$} & \multicolumn{1}{c|}{$0.01$}  & 
\multicolumn{1}{c|}{$0.025$} \\
\hline
\multicolumn{1}{|c|}{$N$ } & \multicolumn{1}{|c|}{$D_{C1}$} & \multicolumn{1}{c|}{$D_{C2}$} & 
\multicolumn{1}{c|}{$D_{C1}$}  & \multicolumn{1}{c|}{$D_{C2}$} & \multicolumn{1}{c|}{$D_{C1}$}  & 
\multicolumn{1}{c|}{$D_{C2}$}\\
\hline
\hline
\multicolumn{7}{|c|}{$D_{SH} function$}\\
\hline
 1000 &   0.01489 &  0.01829  & 0.01761 &  0.02219  &  0.02008 &  0.02407   \\ 
 2000 &   0.01154 &  0.01391  & 0.01309 &  0.01605  &  0.01456 &  0.01778  \\ 
 3000 &   0.01018 &  0.01180  & 0.01116 &  0.01360  &  0.01190 &  0.01462  \\ 
 4000 &   0.00881 &  0.01051  & 0.00955 &  0.01187  &  0.01015 &  0.01273  \\ 
 5000 &   0.00802 &  0.00960  & 0.00869 &  0.01054  &  0.00920 &  0.01120  \\ 
 6000 &   0.00725 &  0.00888  & 0.00797 &  0.00983  &  0.00825 &  0.01014  \\ 
 8000 &   0.00645 &  0.00786  & 0.00680 &  0.00854  &  0.00712 &  0.00911   \\ 
 10000 &  0.00612 &  0.00732  & 0.00641 &  0.00782  &  0.00656 &  0.00827  \\ 
 12000 &  0.00554 &  0.00672  & 0.00575 &  0.00708  &  0.00615 &  0.00761  \\ 
 14000 &  0.00518 &  0.00622  & 0.00541 &  0.00667  &  0.00566 &  0.00714  \\ 
 16000 &  0.00488 &  0.00586  & 0.00518 &  0.00624  &  0.00515 &  0.00658  \\ 
\hline
\hline
\multicolumn{7}{|c|}{$D_{H} function$}\\
\hline
 1000  &   0.01273 &  0.01564  & 0.01733 &  0.02170  & 0.01967 &  0.02372   \\ 
 2000  &   0.01018 &  0.01215  & 0.01258 &  0.01556  & 0.01443 &  0.01742    \\ 
 3000  &   0.00880 &  0.01039  & 0.01063 &  0.01317  & 0.01199 &  0.01449    \\ 
 4000  &   0.00774 &  0.00914  & 0.00907 &  0.01134  & 0.01007 &  0.01264    \\ 
 5000  &   0.00727 &  0.00841  & 0.00827 &  0.01015  & 0.00895 &  0.01103    \\ 
 6000  &   0.00642 &  0.00788  & 0.00774 &  0.00949  & 0.00826 &  0.01007    \\ 
 8000  &   0.00571 &  0.00690  & 0.00668 &  0.00830  & 0.00718 &  0.00905    \\ 
 10000 &   0.00532 &  0.00642  & 0.00622 &  0.00758  & 0.00680 &  0.00825    \\ 
 12000 &   0.00485 &  0.00588  & 0.00566 &  0.00691  & 0.00632 &  0.00765    \\ 
 14000 &   0.00450 &  0.00543  & 0.00529 &  0.00654  & 0.00589 &  0.00714    \\ 
 16000 &   0.00423 &  0.00504  & 0.00501 &  0.00606  & 0.00540 &  0.00665   \\ 
\hline
\hline
\multicolumn{7}{|c|}{$D_D function$}\\
\hline
 1000  &   0.00627 &  0.00750  & 0.00756 &  0.00960  &  0.00807 &  0.00963  \\ 
 2000  &   0.00487 &  0.00583  & 0.00545 &  0.00689  &  0.00581 &  0.00701  \\ 
 3000  &   0.00399 &  0.00485  & 0.00473 &  0.00580  &  0.00480 &  0.00589 \\ 
 4000  &   0.00358 &  0.00434  & 0.00415 &  0.00511  &  0.00427 &  0.00519 \\ 
 5000  &   0.00318 &  0.00394  & 0.00388 &  0.00467  &  0.00376 &  0.00458 \\ 
 6000  &   0.00292 &  0.00356  & 0.00348 &  0.00419  &  0.00347 &  0.00418 \\ 
 8000  &   0.00266 &  0.00321  & 0.00305 &  0.00374  &  0.00312 &  0.00376 \\ 
 10000 &   0.00254 &  0.00302  & 0.00279 &  0.00340  &  0.00279 &  0.00339 \\ 
 12000 &   0.00224 &  0.00275  & 0.00252 &  0.00316  &  0.00264 &  0.00320 \\ 
 14000 &   0.00219 &  0.00261  & 0.00234 &  0.00288  &  0.00251 &  0.00302 \\ 
 16000 &   0.00203 &  0.00242  & 0.00228 &  0.00269  &  0.00230 &  0.00279 \\
\hline
\hline
\end {tabular}
\end {center}
\normalsize
\end {table*}

Also we determined the ratios between the thresholds obtained with different D-functions. For this purpose, from Table \ref{complarge} we used the $D_{C1}$ values corresponding to the probability $P_2=0.01$. The resulting ratios of the thresholds are listed in Table \ref{ratio}.
%
%
\begin{table}[t!]
\small
\caption[] {The average ratios of thresholds $D_{C1}$ taken  from Table \ref{complarge} for fixed $P_2=0.01$ and the size of the orbital sample. In the third columns we added the averaged ratios found for bolides using the small $200$-$800$ synthetic orbits (section \ref{smallsample}).  
}
\label{ratio}
\footnotesize
\begin{center}
\setlength{\extrarowheight}{2pt}
\begin{tabular}{|c|c|c|c|c| }
\hline
\multicolumn{1}{|c|}{D-functions } & \multicolumn{1}{c|}{NEAs} & \multicolumn{1}{c|}{Bolides} & \multicolumn{1}{c|}{Radar}  & \multicolumn{1}{c|}{Video} \\
\hline
 $D_{SH}/D_H$ &   1.14&  1.05  &  1.03 & 0.99 \\ 
 $D_{SH}/D_D$ &   2.44&  2.45  &  2.3  & 2.38 \\ 
\hline
\end {tabular}
\end {center}
\normalsize
\end {table}
One can see that the ratios of the thresholds obtained with $D_{SH}$ and $D_D$ functions are almost the same for all orbital populations used in this study; one can say that it is equal to $\approx 2.4$. However in case of $D_{SH}$ and $D_H$ functions the ratio of the similarity thresholds clearly varies. Usually the $D_{C1}$ thresholds for $D_{SH}$ function are greater than the corresponding values for the $D_H$ function. However this rule fails in case of the SonotaCo 2009 video orbits. In our view, such property shows that despite quite similar forms of $D_{SH}$ and $D_H$  they are not equivalent functions.  

\begin{figure}[b!]
\centerline{
\includegraphics[width=0.5\textwidth]{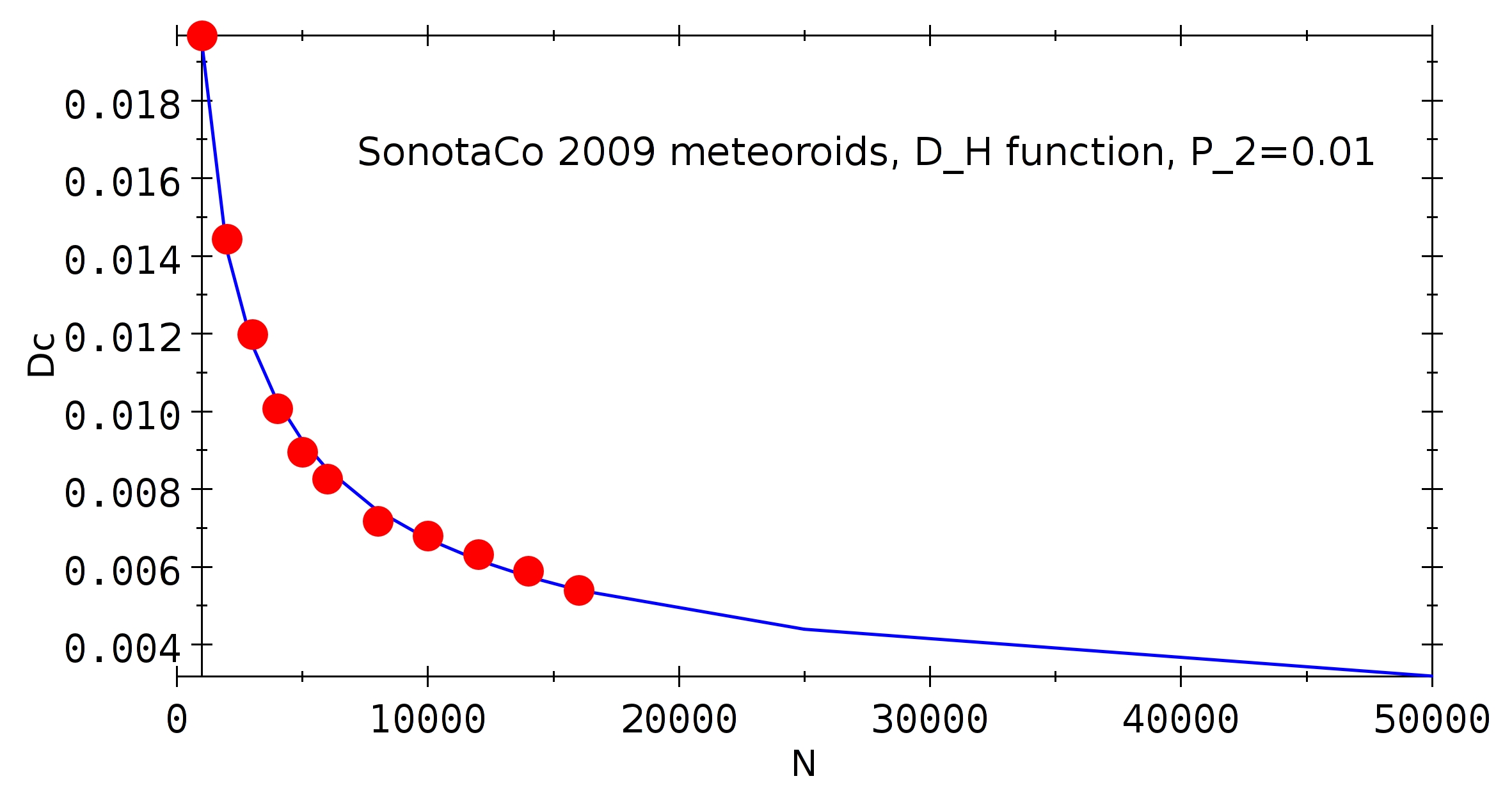}
}
\caption{Thresholds $D_{C}$ versus orbital sample size $N$ for the probability $P_2=0.01$ of a random orbital similarity. The red points correspond to the $N$ and $D_{C1}$ values taken from Table \ref{complarge} for the video samples and the $D_H$ function. The blue curve was obtained by the LSF method. This dependence is given by formula (22).}
\label{fig:DC_N_HY}
\end{figure}

Following the ideas applied in section \ref{smallsample} we have made the LSF fitting using data from Table \ref{complarge}. As result we obtained a series of simple formulae which can be used for calculation of the thresholds of the orbital similarity between two orbits. In Table \ref{formulae} we list the formulae obtained for each D-function and the orbital population used in this study. 
\begin{table}[t!]
\caption[] {The formulae for calculation of the thresholds $D_{C}$  of the orbital similarity for two objects. The thresholds values correspond to the probability of a random similarity $P_2=0.01$.   
}
\label{formulae}
\footnotesize
\begin{center}
\setlength{\extrarowheight}{3pt}
\begin{tabular}[c]{|c|c|cc| }
\hline
\multicolumn{1}{|c|}{ } & \multicolumn{1}{c|}{Population} & \multicolumn{1}{c}{Formula} & \\
\hline
          &      NEAs            &     $D_c=0.2558\cdot N^{-0.408} $  &  (17)\\
 $D_{SH}$ &  Radar Harvard B     &     $D_c=0.4007\cdot N^{-0.450} $  &  (18)\\
          &  Video SonotaCo2009  &     $D_c=0.5837\cdot N^{-0.487} $ &    (19)\\
\hline
          &      NEAs            &     $D_c=0.2193\cdot N^{-0.405} $ &   (20)\\
 $D_{H}$  &  Radar Harvard B     &     $D_c=0.3768\cdot N^{-0.447} $ &   (21)\\
          &  Video SonotaCo2009  &     $D_c=0.4808\cdot N^{-0.464} $ &    (22)\\
\hline
          &      NEAs            &     $D_c=0.1049\cdot N^{-0.408} $ &   (23)\\
 $D_{D}$  &  Radar Harvard B     &     $D_c=0.1543\cdot N^{-0.436} $ &   (24)\\
          &  Video SonotaCo2009  &     $D_c=0.1724\cdot N^{-0.446} $ &   (25)\\
\hline
\end {tabular}
\end {center}
\normalsize
\end {table}
To illustrate the quality of the formulae (17-25) we have chosen formula (22)  for the $D_H$ function and video SonotaCo orbits. On Fig. \ref{fig:DC_N_HY} we plotted  the curve $D_c(N)$ fitted to the SonotaCo 2009 video data taken from Table \ref{complarge}. The red points lie very well on the curve what encouraged us to extend this curve up to $N=50000$. We believe that the formulae (17-25) can be used for the orbital samples of the size of this order.
\section{Conclusions}
In this study we have assessed the probability of a random similarity between two orbits. We tested the influence on this probability of three factors: the choice of the D-function of the orbital similarity, the size of the orbital sample searched for the orbital pairs and the method used for generating the synthetic orbits. All these factors proved to be important. The choice of D-function and the orbital sample size --- their influence is obvious. However, until this study, the influence of the third factor was not known quantitatively. We have shown that, at least in case of the NEAs and the meteoroids populations, their synthetic orbits should not be generated by the method~A --- i.e. assuming the uniform distribution of their orbital elements.In case of the meteoroids the synthetic orbits should be generated by the method~E. 

Also we have shown that in case of searching for similar orbital pairs, one should not used the formula (\ref{row:01}) which give much to high values of the orbital similarity thresholds $D_C$. We found that  in case of the bolide orbital samples the thresholds $D_C$ obtained by formula (\ref{row:01}) correspond to the probability $P_2=1$ of a random similarity. With such $D_C$ values we always find  two ``similar'' objects among the $200$-$800$ orbits or more.

As a remedy for the shortcoming of formula (\ref{row:01}) we proposed its modifications and extensions ($14$--$25$) for two other D-function which one can use to calculate the thresholds $D_C$ corresponding to small random probability $0.01$ of the coincidental similarly between two orbits. They can be used with $D_{SH}$, $D_H$, $D_D$ function and for quite big sets $\sim50000$ objects, we believe) of the NEAs, bolides, radar and video orbits. However at this stage of our study we do not guarantee that these formulae will be as precisely as in case of the samples used in this study, i.e. for all radar or all video meteoroid orbits obtained by different observers. 

In this study we do not exhausted the problem of the assessment of the probability of a random similarity among the orbits of the small bodies. We have left a problem of the orbital pairing with one orbit fixed or the problem of the coincidental grouping of $3, 4, 5...$ orbits. 
We have done some preliminary study of these problems which we intend to study thoroughly in the future. We are aware that these studies are not so different conceptually but much more tough numerically. To continue our study we need much more computer power than we were able to use today.  
\section*{Acknowledgments}
M. Bronikowska work on this paper was supported by the National Science Center (Poland), grant no. 
2013/09/B/ST10/01666.
This research has made use of NASA's Astrophysics Data System Bibliographic Services.
%

\appendix
\section[]{CPD inversion method.}
 The CPD method was elaborated behind astronomy, see e.g. \cite{odwracanie}. It makes use of the inversion of a sample cumulative probability distribution function (CPD). 
The method is based on the transformation of the random variable $\tilde{x}$ into the random variable $\tilde{y}$
\[
{y}=F(x)=\int^x_{-\infty} f(u)du
\]
where $F(x), f(x)$ are the cumulative distribution and the density function, respectively.

\begin{figure}
\centerline{
\includegraphics[width=0.35\textwidth]{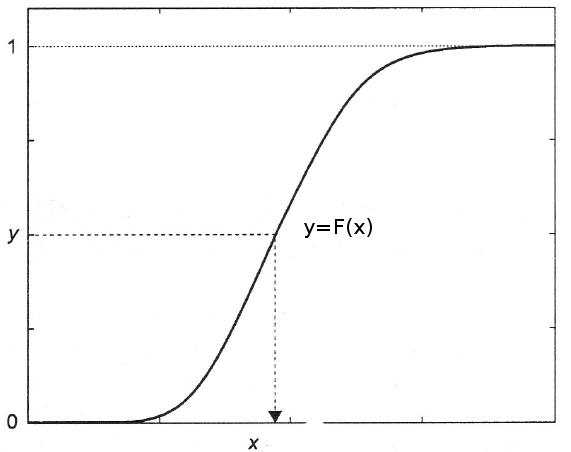}
}
\caption{A random variable $\tilde{y}$ is distributed uniformly $U(0,1)$. Drawing randomly $y$, by means of the inverse function $F^{-1}(y)$, corresponding value of $\tilde{x}$ can be determined.
}
\label{fi:a1}
\end{figure}

The probability density function $g(y)$ of variable $\tilde{y}$ is a constant function, because 
\[
g(y)=f(x=F^{-1}(y))\frac{dx}{dy}=f(x)\left(\frac{dy}{dx}\right )^{-1}=1
\]
This implies that the values of the random variable $\tilde {y}$ one can generate using an uniform random number generator $U(0,1)$. Corresponding values of the second random variable $\tilde x$ one can determine by the inversion of the cumulative distribution function, see Fig. \ref{fi:a1}.   

%
%
In practice we used the cumulative histogram of the random variable $\tilde x$. The value of $x$ was taken as  the value of the corresponding bin. Within a bin, $x$ was varied uniformly.
\section*{References}
\label{lastpage}
\end{document}